\documentclass[a4paper,11pt]{article}
\pdfoutput=1 

\usepackage{jcappub} 

\usepackage[T1]{fontenc} 

\usepackage{graphicx}
\usepackage{dcolumn}
\usepackage{bm}
\usepackage{epsfig}
\usepackage{graphicx}
\usepackage{hyperref}
\usepackage[usenames]{color}
\usepackage{url}
\usepackage{enumitem}
\usepackage{float}
\usepackage{longtable,lscape}

\hypersetup{
    colorlinks=true,
    linkcolor=red,
    citecolor=blue,
}

\newcommand{\rv}{\textcolor{black}}

\newcommand{\remove}[1]{}

\newcommand{\hoallthree}{$  H_0=73.04  \pm  1.04  $ km s$^{-1}$ Mpc$^{-1}$}

\newcommand{\zt}{$z_t= 0.544_{-0.031}^{+0.047}$}

\def\etal{{\frenchspacing\it et al.}}
\def\ie{{\frenchspacing\it i.e.}}

\def\etc{{\frenchspacing\it etc. }}

\def\be{\begin{equation}}
\def\ee{\end{equation}}
\def\ba{\begin{eqnarray}}
\def\ea{\end{eqnarray}}
\title{\boldmath Cosmography via Gaussian Process with Gamma Ray Bursts}

\author{Yuhao Mu,}
\author{Baorong Chang,}
\author[1]{Lixin Xu \note{Corresponding author.}}

\affiliation{Institute of Theoretical Physics, School of Physics, Dalian University of Technology, Dalian, 116024, P. R. China}

\emailAdd{lxxu@dlut.edu.cn}

\abstract{
 
In this paper, we firstly calibrate the Amati relation (the $E_{\rm p}-E_{\rm iso}$ correlation) of gamma ray bursts (GRBs) at low redshifts ($z<0.8$) via Gaussian process by using the type Ia supernovae samples from Pantheon+ under the philosophy that objects at the same redshift should have the same luminosity distance in any cosmology. As a result, this calibration derives the distance moduli of GRBs at high redshifts ($z>0.8$). For an application of these derived distance modulus of GRBs to cosmology, via Gaussian process again, a series of cosmography parameters, which describe kinematics of our Universe, up to the fifth oder and the redshift $z\sim 5$, i.e. the Hubble parameter $H(z)$, the deceleration parameter $q(z)$, the jerk parameter $j(z)$, the snap parameter $s(z)$ and the lerk parameter $l(z)$, are reconstructed from the cosmic observations. The reconstructed cosmography parameters show a transition singularity at $z\sim 6$, it may resort to two possible explanations: one is that the GRBs data points at high redshift $z>5$ are still reliable, it means that new physics beyond the $\Lambda$CDM model happens; another one is that the quality and quantity of GRBs data points at high redshift $z>5$ are not good enough to give any viable prediction of the kinematics of our Universe. To pin down this problem, more high redshifts $z>5$ cosmic observational are still needed.
}

\begin{document}

\maketitle
\flushbottom

\section{Introduction}\label{sec:intro}

Investigating the kinematics of our Universe in a model-independent way is interesting since the discovery of an expanding Universe by E. Hubble in 1929  \cite{Hubble:1929}, now this finding is dubbed as Hubble-Lema\^itre law with memory of Lema\^itre \cite{Lemaitre:2011}. The current expansion rate of our Universe is described by the present Hubble constant $H_0$. However over the last 100 years, the value of $H_0$ was measured by different ways \cite{Freedman:2010xv}, eventually there is still about $5\sigma$ discrepancy of $H_0$ values between the direct and model-independent local measurement $H_0=73.04\pm 1.04 \rm \,km\,s^{-1}\,Mpc^{-1}$ \cite{Riess:2022} from the recent release of the largest type Ia supernovae (SNe Ia) sample called Pantheon+ \cite{Scolnic:2021,Brout:2022} and $H_0=67.4\pm0.5\rm \,km\,s^{-1}\,Mpc^{-1}$ from the Cosmic Microwave Background (CMB) from Planck satellite (PLC18) \cite{Planck2018} in the $\Lambda$CDM cosmology. In order to describe the  kinematics of our Universe, a series of parameters by Taylor expansion of the scale factor $a(t)$ in terms of the cosmic time $t$ are introduced, such as $q$, $j$, $s$, $l$ and so on, named the deceleration, jerk, snap and lerk parameters are defined respectively, for the detailed forms please see Eqs. (\ref{eq:h}, \ref{eq:q}, \ref{eq:j}, \ref{eq:s}, \ref{eq:l}) (see also Eqs. (\ref{eq:hzdc}, \ref{eq:qzdc}, \ref{eq:jzdc}, \ref{eq:szdc}, \ref{eq:lzdc}) in terms of the comoving distance and its derivatives) in the Section \ref{sec:cg}. In the last few years, this kinematics approach has been studied extensively although in different names, for examples cosmography \cite{ref:Turner2002,ref:Visser2004,ref:ST2006,ref:kinematic2,ref:highcosmography,ref:kinematicXu2009,ref:cosmography2011}, cosmokinetics \cite{ref:cosmokinetics,ref:cosmopara}, or Friedmannless cosmology \cite{ref:Friedmannless1,ref:Friedmannless2}. For recent progress, please see Refs. \cite{ref:Bilicki2012,ref:Zhang2016,ref:Haridasu2018,ref:Li2019,ref:Lin2019,ref:Lobo2020,ref:Jesus2022} for instance, but not for a complete list.      
        
In order to investigate the kinematics of our Universe, the distances between galaxies at large scales and their variation with respect to time $t$ (or redshift $z$) are indispensable, just like the findings of the observed galaxies moving away from the Earth at speeds proportional to their distance and the dimmer apparent magnitude of SNe Ia as at high redshifts revealed by \cite{ref:Riess98,ref:Perlmuter99}. Once having the distance indicators along the history of our Universe in hand, one can obtain the kinematics of our Universe. Therefore, the redshift range of distance indicators is demanded as large as possible. As so far, for SNe Ia as standard candles, the observed maximum redshift is $z=2.26137$ \cite{Scolnic:2021,Brout:2022}. And as useful complement, the observed maximum redshift for the gamma ray bursts (GRBs) can reach to $z = 9.4$ \cite{ref:Cucchiara2011}. Although a consensus of GRBs as standard candles is still vanished, several empirical GRBs luminosity relations have been proposed and used in studying cosmology, see \cite{ref:Ghirlanda2006,ref:Schaefer2007,ref:Wang2015,ref:Dainotti2017,ref:Dainotti2018,ref:Amati2013} for reviews. To avoid the circularity problem \cite{ref:Ghirlanda2006} in using GRBs data to constrain cosmological models, one proposes the simultaneous fitting method \cite{ref:Amati2008,ref:Li2008,ref:Wang2008,ref:Xu2012,ref:Amati2013,ref:Khadka2020} and cosmological model-independent method \cite{ref:Liang2008,ref:Liang2008apj} under the assumption that objects at the same redshift should have the same luminosity distance in any cosmology. In GRBs cosmology, the Amati relation \cite{ref:Amati2002,ref:Amati2006}, which is related to the spectral peak energy and the isotropic equivalent radiated energy  (the $E_{\rm p}-E_{\rm iso}$ correlation) of GRBs, is extensively used \cite{ref:Amati2008,ref:Amati2013,ref:Wei2009,ref:Wei2010,ref:Khadka2021,ref:Liang2022}. In the short review paper \cite{ref:Amati2013}, the Amati relation is used to measure the cosmological density parameter $\Omega_m$, where the Amati relation and the cosmological parameter are determined simultaneously. In Ref. \cite{ref:Muccino2021}, the Combo relation \cite{ref:Izzo2015} to a sample of $174$ GRBs is used to investigate possible evidence of evolving dark energy parameter $w(z)$. In Ref. \cite{ref:Liang2022}, Liang \etal ~used the 220 GRB samples (A220) complied by Khadka \etal ~\cite{ref:Khadka2021} to reconstruct the luminosity distance from the Pantheon SNe Ia sample \cite{ref:Scolnic2018} via Gaussian process, where the GRB Hubble diagram at high redshifts was obtained. 

Recently, the largest Supernovae Ia samples was released, dubbed as Pantheon+, which consists of 1701 light curves of 1550 spectroscopically confirmed SNe Ia coming from 18 different sky surveys ranging in redshifts from $z=0.00122$ to $2.26137$ \cite{Scolnic:2021,Brout:2022}. In this paper, we plan to update the distance modulus of GRBs by the recently released Pantheon+ SNe Ia samples mainly following the method proposed in Ref. \cite{ref:Liang2022} but with different redshift range $z\in [0,0.8]$. We obtain $182$ GRBs distance moduli range in the redshifts $0.8<z\le 8.2$. This low redshift SNe Ia calibration seems suspicious that this low redshift SNe Ia method gives very similar results to those obtained via SNe Ia. However, the GRBs at high redshifts where is lack of SNe Ia will give valuable cosmology information. As an instant application in studying cosmology, combining the Pantheon+ SNe Ia samples and the observed Hubble parameters at different redshifts with these derived GRBs distance modulus, we reconstruct the        
kinematics of our Universe in terms of the cosmography parameters $q$, $j$, $s$, $l$ up to the fifth oder and high redshifts.

This paper is organized as follows. In the next Section \ref{sec:cg}, we present the main cosmography parameters. In Section \ref{sec:calibration}, the GRBs Amati relation is calibrated and distance moduli at high redshifts are derived. The cosmography parameter are reconstructed via Gaussian process are given in Section \ref{sec:observation}. The Section \ref{sec:con} is the conclusion.

\section{Cosmography Parameters} \label{sec:cg}

The geometry of our Universe is given by the Friedmann-Lema\^itre-Robertson-Walker (FLRW) metric
\be
ds^2=-c^2dt^2+a^2(t)\left[\frac{dr^2}{1-kr^2}+r^2(d{\theta}^2+\sin^2{\theta}d{\phi}^2)\right],
\ee
where $c$ is the speed of light, $a(t)$ is the scale factor which is normalized to $a_0=1$ at present, $t$ is the cosmic time, $r$ is the comoving coordinate and $\theta$ and $\phi$ are the polar and azimuthal angles in spherical coordinates, the parameter $k=1,0,-1$ denotes three dimensional spatial curvature for closed, flat and open geometries respectively. In this paper, we only consider the spatially flat $k=0$ cosmology.

The cosmography parameters, which describe kinematical state of our Universe and named as Hubble, deceleration, jerk, snap and lerk parameters, are defined as follows respectively,
\ba
H&\equiv&\frac{da(t)}{dt}\frac{1}{a(t)}\equiv\frac{\dot{a}(t)}{a(t)},\label{eq:h}\\
q&\equiv&-\frac{1}{H^2}\frac{d^2a(t)}{dt^2}\frac{1}{a(t)}\equiv-\frac{1}{H^2}\frac{\ddot{a}(t)}{a(t)},\label{eq:q}\\
j&\equiv&\frac{1}{H^3}\frac{d^3a(t)}{dt^3}\frac{1}{a(t)}\equiv\frac{1}{H^3}\frac{a^{(3)}(t)}{a(t)},\label{eq:j}\\
s&\equiv&\frac{1}{H^4}\frac{d^4a(t)}{dt^4}\frac{1}{a(t)}\equiv\frac{1}{H^4}\frac{a^{(4)}(t)}{a(t)},\label{eq:s}\\
l&\equiv&\frac{1}{H^5}\frac{d^5a(t)}{dt^5}\frac{1}{a(t)}\equiv\frac{1}{H^5}\frac{a^{(5)}(t)}{a(t)}.\label{eq:l}
\ea
In term of the redshift $z=1/a(t)-1$, via the relation
\be
\frac{dt}{dz}=-\frac{1}{(1+z)H(z)},
\ee
the cosmography parameters can be rewritten as
\ba
q(z)&\equiv&-1+(1+z)\frac{H'}{H},\label{eq:qz}\\
j(z)&\equiv&1-2(1+z)\frac{H'}{H}+(1+z)^2\frac{H'^2}{H^2}+(1+z)^2\frac{H''}{H},\label{eq:jz}\\
s(z)&\equiv&1-3(1+z)\frac{H'}{H}+3(1+z)^2\frac{H'^2}{H^2}-(1+z)^3\frac{ H'^3}{H^3}  \nonumber\\ 
&-&4(1+z)^3\frac{H'H''}{H^2}+(1+z)^2\frac{H''}{H}-(1+z)^3\frac{H^{(3)}}{H},\label{eq:sz}\\
l(z)&\equiv&1-4(1+z)\frac{H'}{H}+6(1+z)^2\frac{H'^2}{H^2}-4(1+z)^3\frac{ H'^3}{H^3}\nonumber\\ 
&+&(1+z)^4\frac{ H'^4}{H^4}-(1+z)^3\frac{H'H''}{H^2} +7(1+z)^4\frac{H'H'''}{H^2} \nonumber\\ 
&+&11(1+z)^4\frac{H'^2H''}{H^3} +2(1+z)^2\frac{H''}{H}+4(1+z)^4\frac{H''^2}{H^2} \nonumber\\ 
&+&(1+z)^3\frac{H^{(3)}}{H}+(1+z)^4\frac{H^{(4)}}{H},\label{eq:lz}
\ea
where the prime $'$ denotes the derivative with respect to the redshift $z$, and the $f^{(i)}$ denotes the $i$-th order derivative of function $f(z)$ with respect to the redshift $z$.

In order to reconstruct cosmography parameters from cosmic observations, the comoving distances along the line of sight is needed
 \be 
D_C(z)=c\int_0^{z}\frac{dz'}{H(z')}.\label{eq:chi}
\ee  
In terms of $D_C(z)$, the cosmography parameters can be rewritten as
\ba
H(z)&\equiv&\frac{c}{D'_C},\label{eq:hzdc}\\
q(z)&\equiv&-1-(1+z)\frac{ D_C''}{D_C'},\label{eq:qzdc}\\
j(z)&\equiv&\frac{(1+z)^2}{D_C'}\left[\frac{3 D_C''^2}{D_C'}+\frac{2D_C''}{(1+z)}-D_C'''\right],\label{eq:jzdc}\\
s(z)&\equiv&1+ \frac{(1+z)^3 D_C^{(4)}}{D_C'} - \frac{(1+z)^2 D_C^{(3)}}{D_C'} \nonumber\\ 
      &+& \frac{3 (1+z)D_C''}{D_C'}  - \frac{10 (1+z)^3 D_C^{(3)} D_C''}{D_C'^2}   \nonumber\\ 
      &+& \frac{15 (1+z)^3D_C''^3}{D_C'^3} +\frac{5 (1+z)^2 D_C''^2}{D_C'^2} \label{eq:szdc}\\
l(z)&\equiv&1+ \frac{(1+z)^4 D_C^{(5)}}{D_C'} + \frac{(1+z)^3D_C^{(4)}}{D_C'} + \frac{2(1+z)^2D_C^{(3)}}{D_C'}\nonumber\\ 
      &-& \frac{4(1+z) D_C''}{D_C'}  +  \frac{7(1+z)^4 D_C^{(4)}D_C''}{D_C'^2} +\frac{4(1+z)^4 D_C^{(3)2}}{D_C'^2} \nonumber\\   
      &-& \frac{(1+z)^3D_C^{(3)}D_C''}{D_C'^2} +\frac{(1+z)^4 D_C''^4}{D_C'^4} - \frac{4(1+z)^3D_C''^3}{D_C'^3}  \nonumber\\   
      &+&  \frac{11(1+z)^4 D_C^{(3)}D_C''^2}{D_C'^3} +\frac{6(1+z)^2D_C''^2}{D_C'^2} .\label{eq:lzdc}
\ea
It is clear that once the comoving distance and its derivatives are reconstructed, the cosmography parameters and their error bars can be obtained consequently. Here, we would like warning the reader that the Hubble parameter $H(z)$ obviously depends on the present Hubble parameter value $H_0$, but the other cosmography parameters $q(z)$, $j(z)$, $s(z)$ and $l(z)$ are dimensionless and $H_0$ free. The singularity of the cosmography parameters happens when the $D'_c(z)$ crosses zeros at some redshifts.

\section{Calibration to GRBs Amati Relation} \label{sec:calibration}

In this Section, we are mainly going to use the distance moduli from Pantheon+ SNe Ia samples to calibrate the GRBs Amati relation via Gaussian process and then derive distance moduli of GRBs at high redshifts. Therefore we firstly give a brief introduction to the Gaussian process.

Without assuming a specific parameterized form, the Gaussian process can reconstruct the function $f(x)$ from data points $f(x_i)\pm\sigma_i$ via a point-to-point Gaussian distribution \cite{GaPP:2012}. The Gaussian process was used extensively in cosmology study in the last few years \cite{GaPP:2012,Holsclaw:2010prl,Holsclaw:2010prd,Santos-da-Costa:2015,Shafieloo:2012,Yahya:2014,Yang:2015,Cai:2016,Zhang:2016,Cai:2015,Wang:2017}, where the cosmography parameters, equation of state of dark energy are reconstructed by using the cosmic observational data points. The Gaussian process was also used to calibrate GRBs Amati relation in Ref. \cite{ref:Liang2022}. In Gaussian process method, the expected value $\mu$ and the variance $\sigma^2$ of the function $f(x)$ are given by
\ba
\mu(x)&=&\sum_{i,j=1}^Nk(x,x_i)(M^{-1})_{ij}f(x_j),\label{eq:mux}\\
\sigma^2(x)&=&k(x,x)-\sum_{i,j=1}^Nk(x,x_i)(M^{-1})_{ij}k(x_j,x),\label{eq:varx}
\ea
where $N$ is the number of data points. And $M_{ij}=k(x_i,x_j)+C_{ij}$ is the covariance matrix, where $C_{ij}$ is the covariance matrix of the data points, and $k(x, \tilde{x})$ is the covariance function or kernel between the points $x$ and $\tilde{x}$, which is usually taken as the squared exponential covariance function in the form
\be
k(x,\tilde{x}) =
\sigma_f^2 \exp\left[-\frac{(x - \tilde{x})^2}{2\ell^2} \right],\label{eq:kxx}
\ee
where the `hyper-parameter' $\sigma_f$ characterizes the `bumpiness' of the function, i.e. denotes the typical change in the $y$-direction. The length scale
$\ell$  characterizes the distance traveling in $x$-direction to get a significant change in a function. These two `hyper-parameters' $\sigma_f$ and $\ell$ are determined in the Gaussian process by maximizing the logarithmic marginalized likelihood function  
 \be
\ln\mathcal{L}=-\frac{1}{2}\sum_{i,j=1}^Nf(x_i)\left(M^{-1}\right)_{ij}f(x_j)-\frac{1}{2}\ln|M|-\frac{1}{2}N\ln2\pi,
\ee
where $|M|$ is the determinant of $M_{ij}$. In this work, the double squared exponential covariance function 
\be
k(x,\tilde{x}) =
\sigma^2 _{f_{1}}\exp\left[-\frac{(x - \tilde{x})^2}{2\ell_1^2} \right]+\sigma^2 _{f_{2}}\exp\left[-\frac{(x - \tilde{x})^2}{2\ell_2^2} \right],\label{eq:dkxx}
\ee
will also be used to reconstruct the cosmography parameters by considering the GRBs data points at high redshifts and the covariant correlation between them. Fortunately, the above mentioned aspects were already realized in the  {\bf GaPP} code \footnote{\url{https://github.com/carlosandrepaes/GaPP}.} \cite{GaPP:2012}. But, in order to reconstruct $l(z)$, we have modified the {\bf GaPP} code to calculate the fifth order derivative of $D^{(5)}_C$ which is available online \footnote{\url{https://github.com/GraCosPA/Cosmography-GRBs}.}.

\begin{figure*}[tbp]  
\includegraphics[scale=0.9]{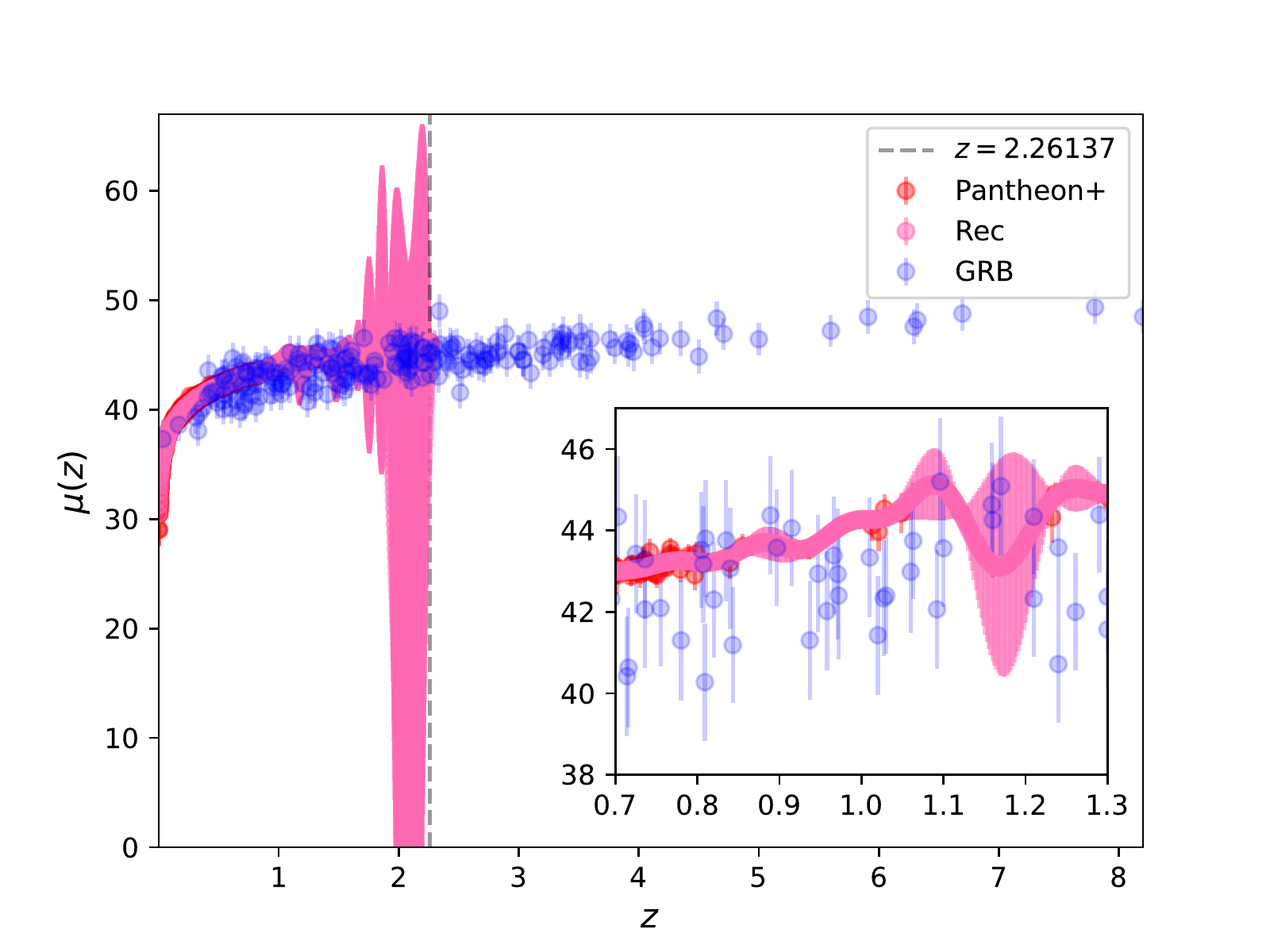}
\caption{The distance moduli of Pantheon+ SNe Ia samples, the reconstructed distance moduli from Pantheon+ SNe Ia samples and the derived distance moduli of GRBs, where the vertical dashed line denotes the maximum redshift of Pantheon+ SNe Ia samples.}\label{fig:mus}
\end{figure*}

For a standard candle such as SNe Ia, the luminosity distance $D_L(z)$ is related to the distance modulus
$\mu=m-M=5\log_{10}D_L({\rm Mpc})+25$, where $M$ is the absolute magnitude of SNe Ia. And the luminosity distance $D_L(z)$, for a spatially flat Universe, is defined as
\be 
D_L(z)=c(1+z)\int_0^{z}\frac{dz'}{H(z')}=(1+z)D_C(z).
\ee 
Thus $D_L=(1+z)D_C$ can be expressed in term of $\mu$ as
\be
D_L=(1+z)D_C=10^{\frac{\mu-25}{5}}{\rm Mpc},
\ee
where $\mu$ is the distance modulus of a SNe Ia, and the absolute magnitude has been determined by the SH0ES Cepheid host distances for Pantheon+ samples \cite{Scolnic:2021,Brout:2022}. It corresponds to set \hoallthree. These moduli of SNe Ia will be used to calibrate GRBs Amati relation under the philosophy that objects at the same redshift should have the same luminosity distance in any cosmology. The distance modulus reconstructed from Pantheon+ SNe Ia samples via Gaussian process with the squared exponential covariance function are shown in Figure \ref{fig:mus} as in pink curves and regions, where the oscillations of the reconstructed function with large error regions are mainly due to sparse data points. It is obvious that these oscillations and large uncertainties are not suitable to calibrate Amati relation. Under this observation, see also the mini figure in Figure \ref{fig:mus}, we prefer calibrating the GRBs Amati relation in the redshift range $z<0.8$ in stated of $z<1.4$ as that did in Ref. \cite{ref:Liang2022}.     

The Amati relation \cite{ref:Amati2002,ref:Amati2006} is given by 
\be
y= a + bx,\label{eq:Amati}
\ee
where $y={\rm log}_{10}\frac{E_{{\rm is o}}}{{\rm 1erg}}$, $x={\rm log}_{10}\frac{E_{{\rm p}}}{{\rm 300keV}}$, $a$ and $b$ are free coefficients to be calibrated by the cosmic observations. Here ${E}_{\rm iso}$ and $E_{\rm p}$ are the isotropic equivalent radiated energy and the spectral peak energy respectively, where $E_{{\rm iso}}$ and $E_{\rm p}$ are related by
\be
E_{{\rm iso}} = 4\pi D^2_L(z)S_{{\rm bolo}}(1+z)^{-1},\quad E_{\rm p} = E^{{\rm obs}}_{{\rm p}}(1+z), \label{eq:GRBsDL}
\ee
where the observables $E^{{\rm obs}}_{{\rm p}}$ and  $S_{{\rm bolo}}$ are the GRBs spectral peak energy and bolometric fluence.

The free coefficients $a$ and $b$ are determined by maximizing the likelihood function
\ba
 \mathcal{L}(\sigma,a,b)\propto\prod_{i=1}^{N} \frac{1}{\sigma}
  \times\exp\left[-\frac{[y_i-y(x_i,z_i; a, b)]^2}{2\sigma^2}\right],\label{Lc}
\ea
where $N=37$ is the number of low-redshift $z<0.8$ GRBs in A220. Here $y_i$ is obtained by the luminosity distance $D_L(z_i)$ reconstructed from SNe Ia data points via Gaussian process 
and the observed $S_{{\rm bolo}}(z_i)$ data point via the Eq. (\ref{eq:GRBsDL}). The $\sigma^2$ is given as \cite{ref:Liang2022}
\be
\sigma^2=\sigma_{\rm int}^2+\sigma_{y,i}^2+b^2\sigma_{x,i}^2,
\ee
where $\sigma_{\rm int}$ is the intrinsic scatter of GRBs, $\sigma_y=\frac{1}{\rm ln10}\frac{\sigma_{E_{\rm iso}}}{E_{\rm iso}},\quad \sigma_x=\frac{1}{\rm ln10}\frac{\sigma_{E_{\rm p}}}{E_{\rm p}}$, $\sigma_{E_{\rm p}}$ is the error magnitude of the spectral peak energy, and $\sigma_{E_{\rm iso}}=4\pi D^2_L\sigma_{S_{\rm bolo}}(1+z)^{-1}$ is the error magnitude of isotropic equivalent radiated energy,
where $\sigma_{S_{\rm bolo}}$ is the error magnitude of bolometric fluence.

\rv{For the Amati relation, actually in the literatures \cite{onoff:Granot2002,jet:Granot2004,onoff:Eichler2004,onoff:Ramirez2005,onoff:Dado2012,onoff:Granot2017,jet:Xu2023}, the on-axis ($\theta_{\rm jet}\ge \theta_{\rm obs}$) and off-axis ($\theta_{\rm jet}< \theta_{\rm obs}$) Amati relation with different indices are discussed for considering the structure of GRBs jets and the effect of the Lorentz factor, where $\theta_{\rm jet}$ is the jet opening angle and $\theta_{\rm obs}$ is the viewing angle. A GRB will become dimmer when the viewing angle is lager than the jet opening angle because of the relativistic beaming effect \cite{onoff:Granot2002,jet:Huang2002,jet:Yamazaki2003}, therefore the derived luminosity distance of a GRB would be affected by its jet structure. In consequence the inferred cosmological model parameters would depend on the structure of GRBs jets. In the recent paper \cite{jet:Xu2023}, the empirical correlations of GRBs in the on-axis and off-axis cases with and without the effect of the Lorentz factor are derived analytically, where $E_{\rm p} \propto E^{0.5}_{\rm iso}$ for the on-axis case and $E_{\rm p} \propto E^{1/4\sim 4/13}_{\rm iso}$ for the off-axis case are obtained respectively. In fact a general index, \ie~$E_{\rm p} \propto E^{1/b}_{\rm iso}$ as seen from the Eq. (3.8), is considered in this work, where the information about the structure of GRBs jets, the effect of the Lorentz factor and other possible unknown effects are equivalently included roughly.}

\begin{figure*}[tbp]  
\includegraphics[scale=0.7]{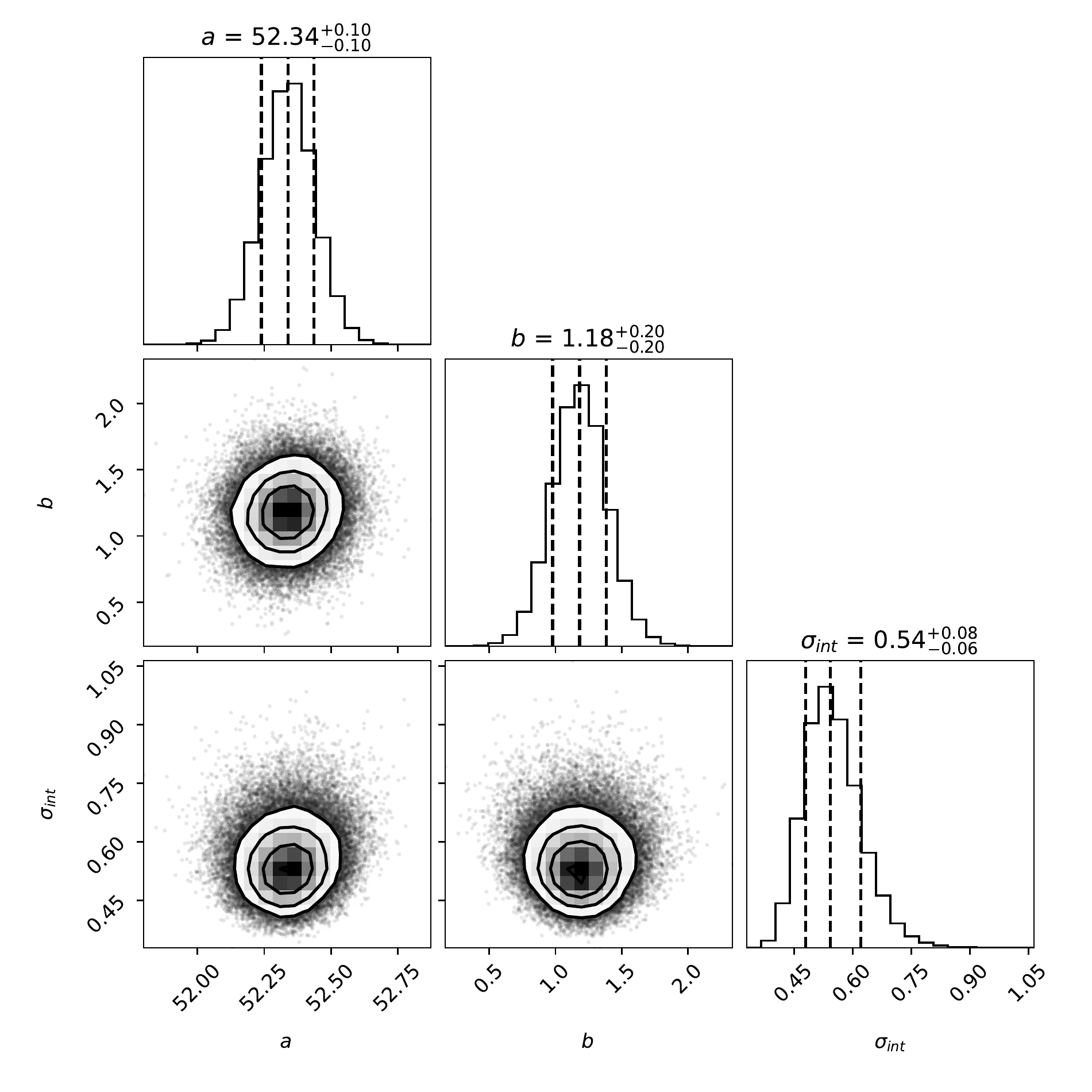}
\caption{Contour plots for the Amati relation coefficients and the intrinsic scatter of GRBs, where the redshifts of GRBs in the range of $z<0.8$ are used.}\label{fig:corner}
\end{figure*}

It is clear that GRBs can not be calibrated if the absolute magnitude $M$ of SNe Ia is still not known, even one takes $\mu+M$, i.e. the apparent magnitude $m$, as observable. It implies the degeneracy between the Amati relation parameter $a$ and the absolute magnitude $M$, and only the Amati relation parameter $b$ can be constrained. Implementing Markov Chain Monte Carlo numerical fitting method by using GRBs data points ranging in $z<0.8$, one obtains the Amati relation with fixed coefficients $a=52.34\pm 0.10$, $b=1.18\pm 0.20$ and $\sigma_{\rm int}=0.54^{+0.08}_{-0.06}$ \footnote{Actually, by using GRBs at $z<1.4$ and repeating the process, one has $a=52.30\pm 0.07$, $b=1.06\pm 0.12$ and $\sigma_{\rm int}=0.51^{+0.05}_{-0.04}$.}. The corresponding contour is plotted in Figure \ref{fig:corner}.
Once the Amati relation was calibrated at low redshifts, the distance modulus at high redshift will be obtained easily from Eq. (\ref{eq:GRBsDL}), and the corresponding uncertainty of the GRBs distance modulus is given by \cite{ref:Liang2022}
\be
\sigma^2_\mu=\bigg(\frac{5}{2}\sigma_{{\rm log}\frac{E_{\rm iso}}{\rm 1erg}}\bigg)^2+ \bigg (\frac{5}{\rm 2ln10}\frac{\sigma_{S_{\rm bolo}}}{S_{\rm bolo}} \bigg)^2,
\ee
where
\ba
\sigma^2_{{\rm log}\frac{E_{\rm iso}}{\rm 1erg}}&=&\sigma^2_{\rm int}+ \bigg (\frac{b}{\rm ln10}\frac{\sigma_{E_{\rm p}}}{E_{\rm p}} \bigg )^2\nonumber\\
&+&\sum_{ij} \left[\frac{\partial {y}(x;\theta_c)}{\partial \theta_i} \right]C_{ij}\left[\frac{\partial {y}(x;\theta_c)}{\partial \theta_j}\right], 
\ea
where $\theta_c$=\{$\sigma_{{\rm int}}$, $a$, $b$\}, and $C_{ij}$ is the covariance matrix of these fitting coefficients. For convenience, the $182$ derived distance moduli for GRBs at redshift $z>0.8$ are summarized in the appendix \ref{sec:appendix}. Now these derived GRBs distance moduli can be used to constrain cosmology models and properties of dark energy. In particular, these distance moduli are compatible to Pantheon+ SNe Ia samples and can be used simultaneously under the philosophy that objects at the same redshift should have the same luminosity distance in any cosmology. Meanwhile, we should mention that the GRB051109A sample is removed as did in Ref. \cite{ref:Liang2022} for different values reported in Refs. \cite{ref:Amati2008} and \cite{ref:Demianski2017a}.

\begin{figure*}[tbp]  
\includegraphics[scale=0.5]{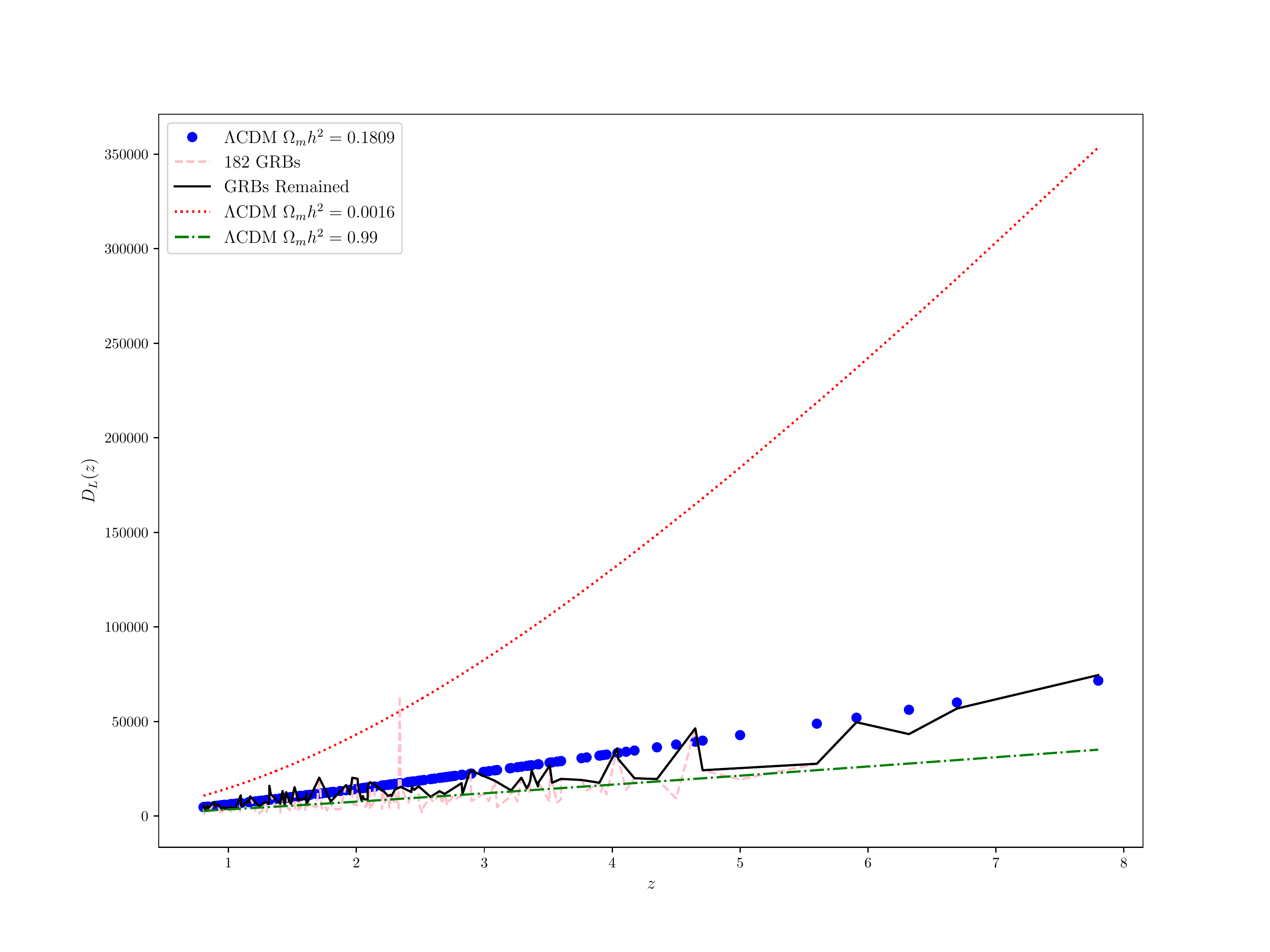}
\caption{The luminosity distance $D_L(z)$ for the $\Lambda$CDM model with different values of $\Omega_m h^2$ and calibrated GRBs, where the blue circles denote the luminosity distance predicted by $\Lambda$CDM model with $\Omega_m h^2=0.1809$, and the black solid line denotes the remained GRBs after physically reasonable and meaningful selection.}\label{fig:selectedDL}
\end{figure*}

Before using these $182$ derived GRBs distance moduli to do cosmological constraints or cosmography parameters reconstruction, one should make sure that all these data points are physically reasonable and meaningful. It is obvious that a physically reasonable comoving distance $D_C(z)$ should increase with the redshift at the late epoch of our Universe. It implies that $H(z)=c/D'_C(z)$ is alway positive and singularity free. Keeping this in mind, one assumes that the physically meaningful luminosity distance of GRBs should lie in the range predicted by the standard $\Lambda$CDM model with different $\Omega_m h^2$ values ranging in $[0.0016, 0.99]$, which corresponds to $\Omega_m=0.01$, $h=0.40$ and  $\Omega_m=0.99$, $h=1.00$ respectively. The $\Omega_m h^2$ ranges largely in almost all physically reasonable cosmology. Although this simple assumption might not keep $D_C(z)$ alway increase with the redshift, physically reasonable and meaningful data points are selected. In this way, one finally selects $97$ data points from $182$ GRBs, where the higher redshift $z=8.2$ data point was removed due to its apparent depressed central value. Meanwhile the GRB050904 at $z=6.29$ is also removed, due to its smaller central value with comparison to GRB140515A at $z=6.32$. The final data selection results can be seen in Figure \ref{fig:selectedDL}, where the blue circles denote the luminosity distance predicted by $\Lambda$CDM model with $\Omega_m h^2=0.1809$, and the black solid line denotes the remained GRBs after the physically reasonable and meaningful selection.

\section{Reconstructed Cosmography Parameters via the Gaussian Process} \label{sec:observation}

As a direct application to cosmology, we move to study the kinematics of our Universe based on the selected observational data points. For seeking that, we resort to using the Gaussian process again, but with the double squared exponential covariance function given by Eq. (\ref{eq:dkxx}). This consideration is based on the fact that GRBs as contrast to Pantheon+ samples ranges in a large redshift range and has large distance modulus uncertainty. The extra hyper-parameters $\sigma_f$ and $\ell$ can handle this diversity. In fact, we have tasted and confirmed that the squared exponential covariance function really gives weird oscillations, but the double squared exponential covariance function will not.           

In order to reconstruct $D_C$ and its derivatives by using the Gaussian process code {\bf GaPP} \cite{GaPP:2012}, the covariance matrix for the new observable $D_C=D_L/(1+z)$, which can be derived by error propagation equation, is given as 
\be
C^{\rm tot}_{ij} =\left[\frac{D^i_L}{(1+z_i)^2}\right]^2\sigma^2_{z_{i}}\delta_{ij}+\frac{\ln10 D^i_L}{5(1+z_i)} \tilde{C}^{\rm tot}_{ij}\frac{\ln10 D^j_L}{5(1+z_j)} , \label{eq:covariance}
\ee 
where $z_i$ and $D^i_L$ are the redshift and the observed luminosity distance of the $i$-th SN Ia respectively, and $\sigma_{z_{i}}$ is the $1\sigma$ error for $z_i$. And $\delta_{ij}$ is the standard Kronecker symbol. $\tilde{C}^{\rm tot}_{ij}$ in the last term is total distance covariance matrix for Pantheon+ SN Ia samples \footnote{The data points are available online \url{https://github.com/PantheonPlusSH0ES/DataRelease}.} \cite{Scolnic:2021,Brout:2022}, and there is no Einstein's summation convention. This variance $C^{\rm tot}_{ij}$ has to be added to the covariance matrix
\be
{\bm y} \sim \mathcal{N}\left( \text{\boldmath $\mu$},K(\bm X,\bm X) + C^{\rm tot}\right),
\ee
where $[K(\bm X,\bm X)]_{ij}=k(x_i,x_j)$ is the covariance matrix for  a set of input points $\bm X=\{x_i\}$. Similarly, in order to reconstruct $D_C'$ from the cosmic chronometers (CC), the following covariance matrix is needed
\be
C^{\rm H}_{ij} = \left[\frac{c}{H_i^2}\right]^2\sigma^2_{H_i} \delta_{ij}. \label{eq:Hcovariance}
\ee  
Here the squared exponential covariance function Eq. (\ref{eq:dkxx}) is taken as the covariance function, which is also infinitely differentiable and useful for reconstructing the derivative of a function. 

The recent release of the Pantheon+ samples contains SN Ia ranging in redshifts from $z=0.00122$ to $2.26137$, which consists of 1701 light curves of 1550 spectroscopically confirmed SN Ia coming from 18 different sky surveys. As pointed as in our previous study \cite{ref:cosmography2011}, due to the degeneracy between $H_0$ and the absolute magnitude $M$, the SN Ia cannot give any prediction of $H_0$ value without calibration. Therefore, in this work, we use $H_0$ from SH0ES to reconstruct $H(z)$. In using the measurement of $H_0$ from SH0ES, and making it consistent and free of redundancy, some Pantheon+ SN Ia data points (marked as {\bf USED\_IN\_SH0ES\_HF=1}) are removed where they were already used in the Hubble flow dataset \cite{Riess:2022}.  

For the observational Hubble data, or the so-called cosmic chronometers (CC) which is determined by computing the age difference $\Delta t$ between passively-evolving galaxies at close redshifts, the sample compiled by \cite{Moresco2022CCdata} is used, see also the data table available online \footnote{\url{https://github.com/carlosandrepaes/GaPP}.}, where the redshift ranges in $z\in[0.070, 2.360]$.

\begin{figure*}[tbp]  
\includegraphics[scale=0.5]{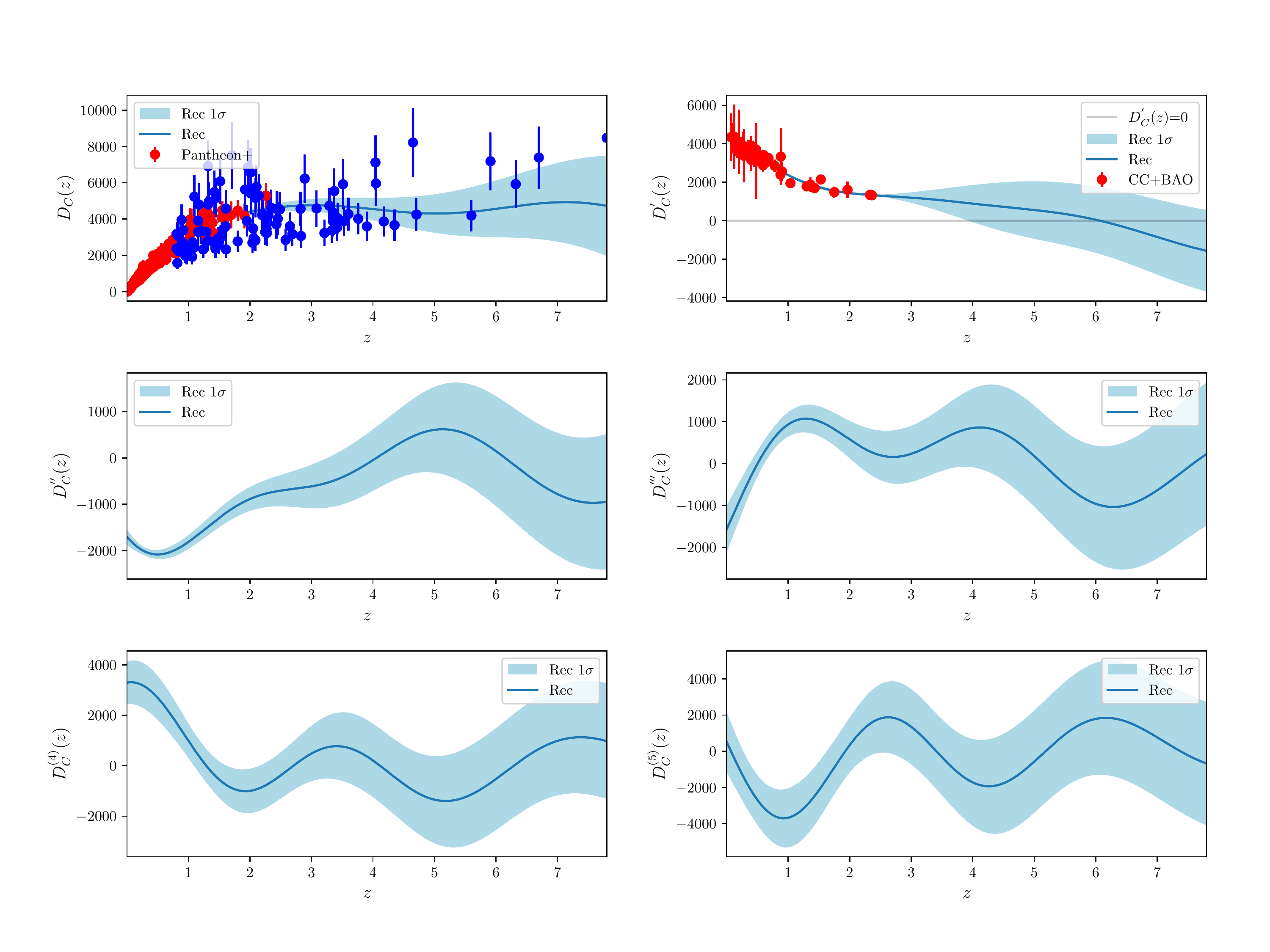}
\caption{The reconstructed cosmography parameters $D_C(z)$, $D'_C(z)$, $D''_C(z)$, $D'''_C(z)$, $D^{(4)}_C(z)$ and $D^{(5)}_C(z)$ (with $1\sigma$ error region) with the joint CC, Pantheon+ SN Ia samples and high redshift GRBs from the upper left panel to the lower right panel respectively.}\label{fig:obsdcs}
\end{figure*}

Implementing the Gaussian process as described in the Section \ref{sec:calibration}, the comoving distance and its derivatives up to the fifth oder with respect to the redshift are reconstructed as shown in Figure \ref{fig:obsdcs}, where $1\sigma$ errors are also plotted in shadow regions. It is seen that the error becomes larger with the increase of the oder of derivative with respect to the redshift $z$. On the contrary, the addition of CC data points gives an extra constraint to the first order derivative of $D_C(z)$, thus a relative narrow error region for the reconstructed functions can be obtained. Meanwhile, a large error is shown at high redshift due to the sparse data points at where.       

\begin{figure*}[tbp]  
\includegraphics[scale=0.5]{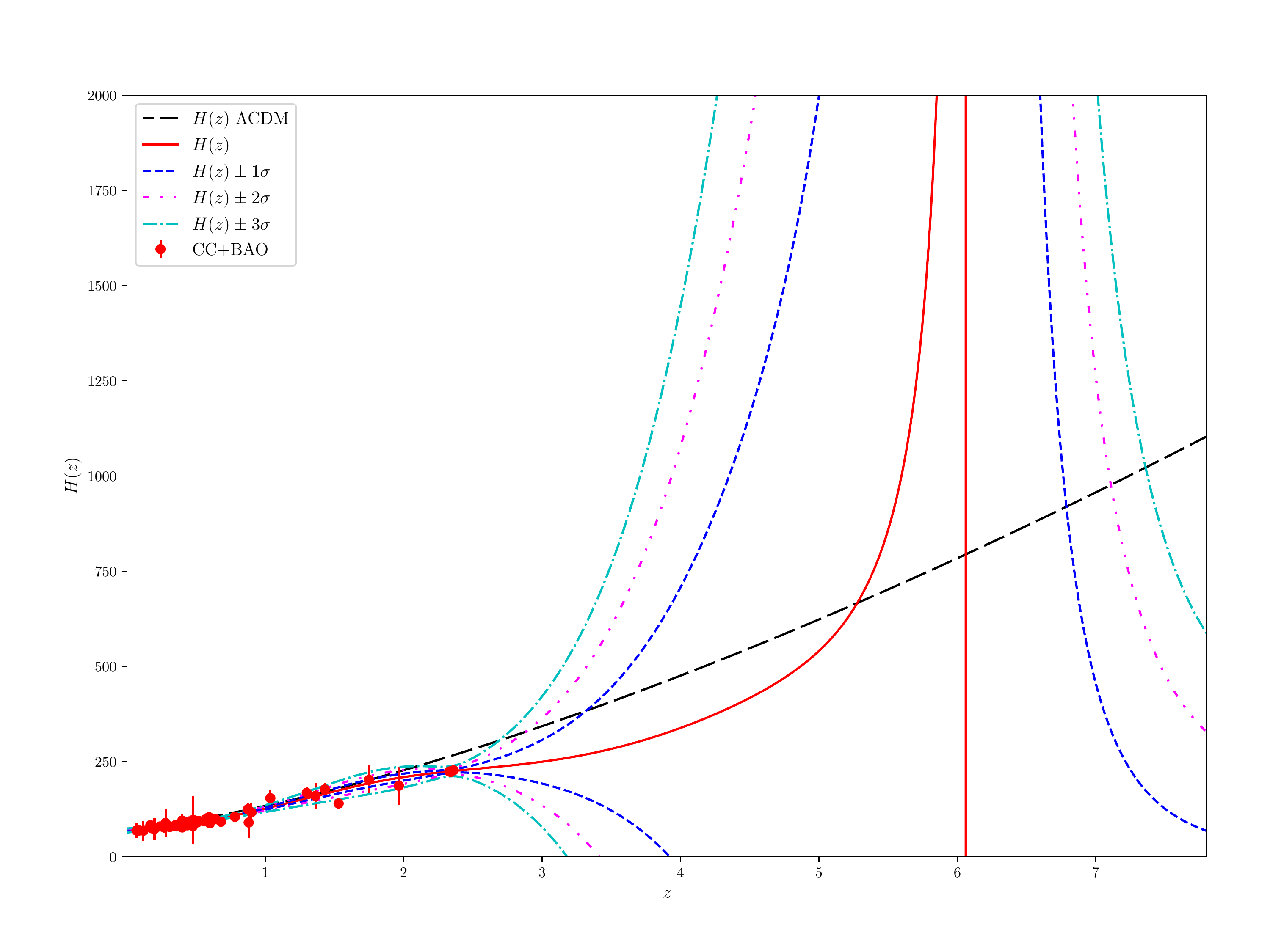}
\caption{The reconstructed Hubble parameter $H(z)$ (with $1-3\sigma$ error regions) with the joint CC and Pantheon+ SN Ia samples, where the Hubble parameter predicted from a spatially flat $\Lambda$CDM model is also plotted as for comparison.}\label{fig:obshz}
\end{figure*}

\begin{figure*}[tbp]  
\includegraphics[scale=0.5]{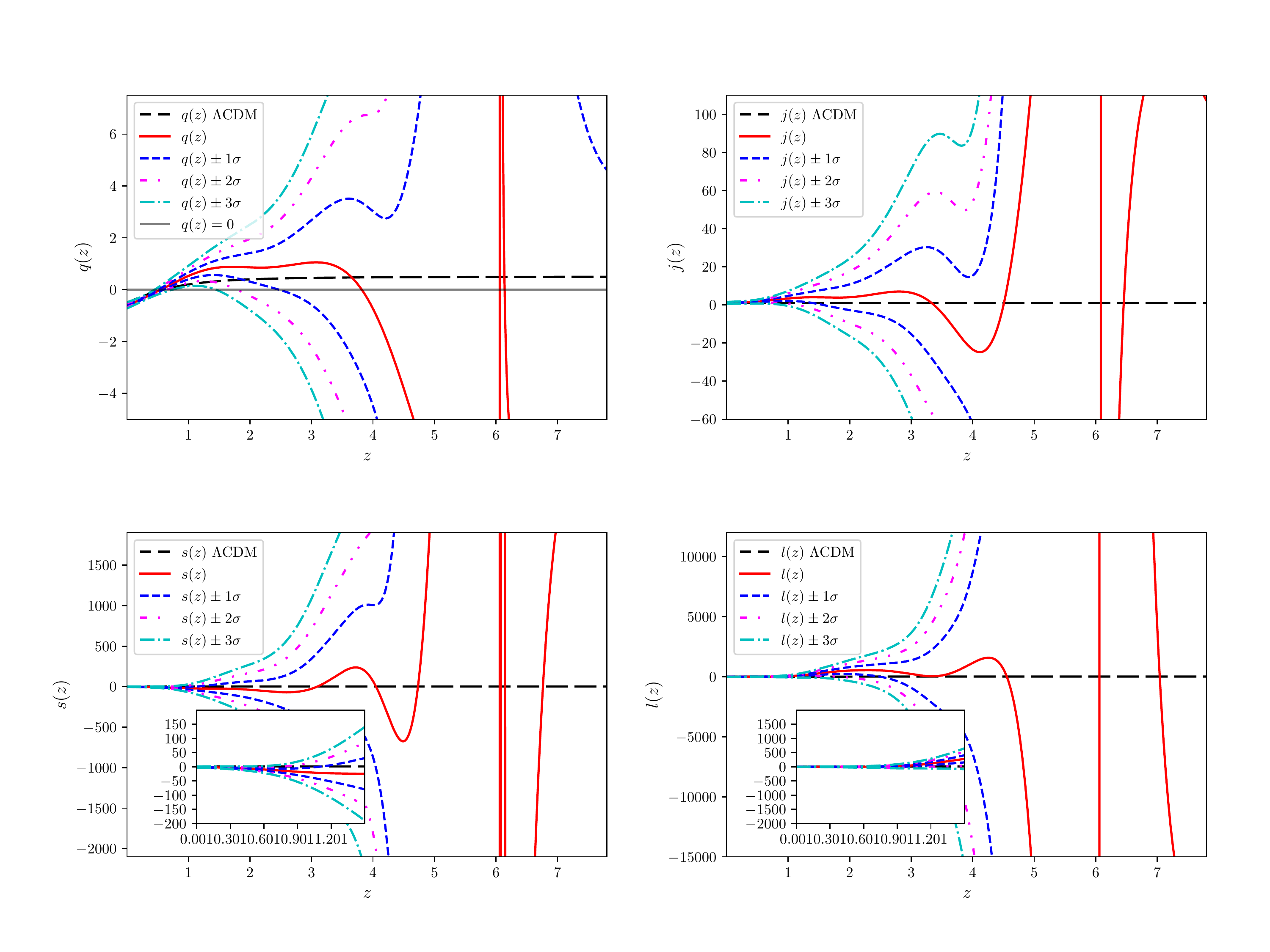}
\caption{The reconstructed cosmography parameters $q(z)$, $j(z)$, $s(z)$ and $l(z)$ (with $1-3\sigma$ error curves) with the joint CC, Pantheon+ SN Ia samples and high redshift GRBs from the upper left panel to the lower right panel respectively, where the corresponding cosmography parameters predicted from a spatially flat $\Lambda$CDM model are also plotted as for comparison. In the upper right $q(z)$ panel, the horizon $q(z)=0$ line is for showing the transition redshift (at \zt) from a decelerated expansion to an accelerated expansion at the crossing point with the reconstructed $q(z)$ red solid line.}\label{fig:obscosmography}
\end{figure*}

With the joint CC and Pantheon+ SN Ia samples, the reconstructed Hubble parameter $H(z)$ is shown in Figure \ref{fig:obshz} including $1-3\sigma$ error curves, where the Hubble parameter $H(z)$ predicted from a spatially flat $\Lambda$CDM cosmology, i.e. $H^2(z)=H^2_0[\Omega_{m0}(1+z)^3+\Omega_{\Lambda0}]$ with $\Omega_{m0}=0.334$  ($\Omega_{\Lambda0}=1-\Omega_{m0}$) from SH0ES \cite{Riess:2022} is also plotted as for comparison. The apparent bumps of error curves for $H(z)$ at the redshift range $z\sim1.0-2.0$ are mainly due to the sparse and large error bars of the data sets. The vertical lines in Figure \ref{fig:obshz} happen at the redshifts where $D'_{C}(z)$ cross zero line, i.e. when the comoving distance $D_C(z)$ transfers from increase to decrease or inverse with respect to the redshift $z$. The same situation appears in the reconstructed cosmography parameters $q(z)$, $j(z)$, $s(z)$ and $l(z)$ as shown in Figure  \ref{fig:obscosmography}, where the corresponding cosmography parameters predicted from the spatially flat $\Lambda$CDM cosmology are also plotted as for comparison. The corresponding error is obtained by the error propagation equation, say for a function looks like $f=g^m/h^n$, the errors, after omitting the cross correlation between $g$ and $h$, can be calculated as
\be
\sigma^2_f=\left[\frac{ng^m}{h^{n+1}}\right]^2\sigma^2_h + \left[\frac{mg^{m-1}}{h^n}\right]^2\sigma^2_g.
\ee
Thus the corresponding calculation for $\sigma_q$ \etc is quite easy, but the mathematical expression is long and ugly, so it is not shown in this paper. 

In the upper left $q(z)$ panel of Figure \ref{fig:obscosmography}, the horizon $q(z)=0$ line is for showing the transition redshift (at \zt) from a decelerated expansion to an accelerated expansion at the crossing point with the reconstructed $q(z)$ red solid line. This transition redshift is lower than that $z_t\sim 0.586$ predicted by the spatially flat $\Lambda$CDM model. The evolution of the reconstructed cosmography parameters $q(z)$, $j(z)$, $s(z)$ and $l(z)$ with respect to the redshift $z$ becomes weird at high redshifts $z>5$. This strange behavior can simply boil down to the definition of cosmography parameters in terms of $D'_C(z)$ which appears in the denominator of the corresponding expression. The singularities happen when $D'_C(z)$ crosses zeros. Of course this singularity means a transition of the comoving distance. But this transition seems unphysical in a regular cosmology. At least at the late epoch of our Universe, the luminosity distance and comoving distance should increase with the redshift for a reasonable cosmological model, thus the transition singularity for $H(z)=c/D'_{C}(z)$, i.e. $D'_C(z)=0$, cannot be encountered. Therefore, one has two possible explanations to this transition singularity: one is that the GRBs data points at high redshift $z>5$ are still reliable, it means that new physics beyond the $\Lambda$CDM model happens; another one is that the quality and quantity of GRBs data points at high redshift $z>5$ are not good enough to give any viable prediction of the kinematics of our Universe. To pin down this problem, more high redshifts $z>5$ cosmic observational are still needed.

Since we have already reconstructed the evolutions of cosmography parameters within $1-3\sigma$ regions with respect to the redshift ranging in $z\in [0,8]$, the current values of the cosmography parameters as byproducts can be easily read: $q_0=-0.608\pm0.040$, $j_0=1.038\pm0.175$, $s_0=-0.608\pm0.668$ and $l_0=1.596\pm0.930$.

\section{Conclusion} \label{sec:con}

In this paper, the GRBs Amati relation is calibrated by using Gaussian process from Pantheon+ SN Ia samples at low redshifts. After doing that, we obtain the $182$ GRBs distance moduli in the redshift ranging in $z>0.8$. Usually these derived GRBs distance moduli can be used to constrain cosmology and dark energy properties, but before doing that one has to consider the physically reasonable and meaningful data points to give reliable predictions. To select the physically reasonable and meaningful data points, one assumes the meaningful luminosity distance of GRBs should lie in the range predicted by the standard $\Lambda$CDM model with different $\Omega_m h^2$ values ranging in $[0.0016, 0.99]$, which corresponds to $\Omega_m=0.01$, $h=0.40$ and  $\Omega_m=0.99$, $h=1.00$ respectively. The $\Omega_m h^2$ range is chosen to cover almost all physically reasonable cosmology in a spatially flat $\Lambda$CDM model. Meanwhile a smaller central value data point is also removed when two GRBs are very close, thus the GRB050904 at $z=6.29$ is also removed. This selection reduces the data points of GRBs from $182$ to $96$. As a direct application to cosmology for these selected GRBs distance moduli, the cosmography parameters up to the fifth order are reconstructed by combining the cosmic observational data pints from Pantheon+ SN Ia samples, CC and GRBs at high redshifts. It shows that the reconstructed cosmography parameters are compatible with with that from $\Lambda$CDM in $3\sigma$ regions. These reconstructed cosmography parameters evolve well with respect to the redshift $z$, until they encounter transition singularities at $z\sim 6$ as shown by the vertical lines in Figure \ref{fig:obshz} and Figure \ref{fig:obscosmography}. These singularities happen when $D'_{C}(z)$ crosses zero as shown in the upper right panel of Figure \ref{fig:obsdcs}. This crossing makes the Hubble parameter goes to infinity via the relation $H(z)=c/D'_{C}(z)$. Of course, this singularity means the comoving distance evolves with the redshift from increasing to decreasing, or vice versa. However in any case, this transition seems unphysical in a regular cosmology at the late epoch. Apparently one has two possible explanations to this transition singularity: one is that the GRBs data points at high redshift $z>5$ are still reliable, it means that new physics beyond the $\Lambda$CDM model happens; another one is that the quality and quantity of GRBs data points at high redshift $z>5$ are not good enough to give any viable prediction of the kinematics of our Universe. To pin down this problem, more high redshifts $z>5$ cosmic observational are still needed. Leaves it alone, we have reconstructed the cosmography parameters up to $z\sim 5$ and the fifth oder. This is main findings of this work. And we hope this study may shed some lights on the kinematics of our Universe. 

\acknowledgments

We are grateful to the anonymous referee who made many insightful comments and suggestions that are very helpful for improving the quality of this work. This work is supported in part by National Natural Science Foundation of China under Grant No. 12075042 and No. 11675032.

\begin{appendix} 
\section{The GRB data sets}\label{sec:appendix}

\begin{longtable} {lcc}
\caption{List of the Derived Distance Moduli of $182$ GRBs in the A220 Sample at redshift $0.8<z\le 8.2$. The GRB051109A sample is removed as did in Ref. \cite{ref:Liang2022} for different values reported in Refs. \cite{ref:Amati2008} and \cite{ref:Demianski2017a}.}\\
\hline\hline
GRB &   $z$   & $\mu_{\rm GRB}\pm \sigma_{\mu,\rm GRB }$  \\
\hline
\endfirsthead
\caption{continued.}\\
\hline\hline
GRB &   $z$   & $\mu_{\rm GRB}\pm \sigma_{\mu,\rm GRB }$  \\
\hline
\endhead
\hline
\endfoot
100816A   & $ 0.8049 $  &  $ 43.52  \pm   1.42 $ \\
150514A   & $ 0.8070 $  &  $ 43.17  \pm   1.43 $ \\
051022   & $ 0.8090 $  &  $ 40.28  \pm   1.44 $ \\
151027A   & $ 0.8100 $  &  $ 43.80  \pm   1.45 $ \\
110715   & $ 0.8200 $  &  $ 42.30  \pm   1.42 $ \\
970508   & $ 0.8350 $  &  $ 43.76  \pm   1.49 $ \\
060814   & $ 0.8400 $  &  $ 43.07  \pm   1.49 $ \\
990705   & $ 0.8430 $  &  $ 41.19  \pm   1.43 $ \\
140506A   & $ 0.8890 $  &  $ 44.37  \pm   1.46 $ \\
091003A   & $ 0.8969 $  &  $ 43.58  \pm   1.44 $ \\
141225A   & $ 0.9150 $  &  $ 44.06  \pm   1.43 $ \\
080319B   & $ 0.9370 $  &  $ 41.30  \pm   1.46 $ \\
071010B   & $ 0.9470 $  &  $ 42.94  \pm   1.44 $ \\
970828   & $ 0.9580 $  &  $ 42.02  \pm   1.46 $ \\
980703   & $ 0.9660 $  &  $ 43.39  \pm   1.45 $ \\
091018   & $ 0.9710 $  &  $ 42.93  \pm   1.61 $ \\
160131A   & $ 0.9720 $  &  $ 42.40  \pm   1.56 $ \\
021211   & $ 1.0100 $  &  $ 43.34  \pm   1.47 $ \\
991216   & $ 1.0200 $  &  $ 41.43  \pm   1.46 $ \\
140508A   & $ 1.0270 $  &  $ 42.34  \pm   1.42 $ \\
080411   & $ 1.0300 $  &  $ 42.40  \pm   1.43 $ \\
000911   & $ 1.0600 $  &  $ 42.99  \pm   1.51 $ \\
091208B   & $ 1.0630 $  &  $ 43.75  \pm   1.42 $ \\
091024   & $ 1.0920 $  &  $ 42.06  \pm   1.46 $ \\
980613   & $ 1.0960 $  &  $ 45.20  \pm   1.56 $ \\
080413B   & $ 1.1000 $  &  $ 43.57  \pm   1.44 $ \\
061126   & $ 1.1588 $  &  $ 44.63  \pm   1.52 $ \\
130701A   & $ 1.1600 $  &  $ 44.26  \pm   1.42 $ \\
160509A   & $ 1.1700 $  &  $ 45.09  \pm   1.71 $ \\
140213A   & $ 1.2100 $  &  $ 42.33  \pm   1.42 $ \\
140907A   & $ 1.2100 $  &  $ 44.34  \pm   1.42 $ \\
090926B   & $ 1.2400 $  &  $ 43.58  \pm   1.42 $ \\
130907A   & $ 1.2400 $  &  $ 40.72  \pm   1.44 $ \\
061007   & $ 1.2610 $  &  $ 42.00  \pm   1.44 $ \\
131030A   & $ 1.2900 $  &  $ 44.39  \pm   1.42 $ \\
990506   & $ 1.3000 $  &  $ 41.57  \pm   1.46 $ \\
130420A   & $ 1.3000 $  &  $ 42.37  \pm   1.43 $ \\
061121   & $ 1.3140 $  &  $ 44.07  \pm   1.47 $ \\
141220A   & $ 1.3195 $  &  $ 45.23  \pm   1.42 $ \\
140801A   & $ 1.3200 $  &  $ 46.03  \pm   1.42 $ \\
071117   & $ 1.3310 $  &  $ 45.32  \pm   1.50 $ \\
100414A   & $ 1.3680 $  &  $ 43.35  \pm   1.46 $ \\
120711A   & $ 1.4050 $  &  $ 44.08  \pm   1.51 $ \\
160625B   & $ 1.4060 $  &  $ 41.40  \pm   1.47 $ \\
151029A   & $ 1.4230 $  &  $ 45.62  \pm   1.49 $ \\
100814   & $ 1.4400 $  &  $ 43.81  \pm   1.44 $ \\
050318   & $ 1.4400 $  &  $ 43.83  \pm   1.46 $ \\
141221A   & $ 1.4520 $  &  $ 45.51  \pm   1.44 $ \\
110213   & $ 1.4600 $  &  $ 43.28  \pm   1.48 $ \\
010222   & $ 1.4800 $  &  $ 42.43  \pm   1.44 $ \\
120724   & $ 1.4800 $  &  $ 44.30  \pm   1.50 $ \\
060418   & $ 1.4890 $  &  $ 44.07  \pm   1.48 $ \\
150301B   & $ 1.5169 $  &  $ 45.92  \pm   1.44 $ \\
030328   & $ 1.5200 $  &  $ 42.25  \pm   1.43 $ \\
070125   & $ 1.5470 $  &  $ 42.81  \pm   1.46 $ \\
090102   & $ 1.5470 $  &  $ 44.54  \pm   1.48 $ \\
161117A   & $ 1.5490 $  &  $ 41.76  \pm   1.42 $ \\
060306   & $ 1.5590 $  &  $ 44.71  \pm   1.54 $ \\
040912   & $ 1.5630 $  &  $ 43.41  \pm   1.79 $ \\
100728A   & $ 1.5670 $  &  $ 42.76  \pm   1.43 $ \\
990123   & $ 1.6000 $  &  $ 42.55  \pm   1.52 $ \\
071003   & $ 1.6040 $  &  $ 44.86  \pm   1.50 $ \\
090418   & $ 1.6080 $  &  $ 45.39  \pm   1.53 $ \\
110503   & $ 1.6100 $  &  $ 43.92  \pm   1.43 $ \\
990510   & $ 1.6190 $  &  $ 43.60  \pm   1.43 $ \\
080605   & $ 1.6398 $  &  $ 43.87  \pm   1.43 $ \\
131105A   & $ 1.6900 $  &  $ 43.31  \pm   1.44 $ \\
091020   & $ 1.7100 $  &  $ 46.54  \pm   1.69 $ \\
120119   & $ 1.7300 $  &  $ 43.00  \pm   1.43 $ \\
100906   & $ 1.7300 $  &  $ 43.19  \pm   1.64 $ \\
150314A   & $ 1.7580 $  &  $ 43.37  \pm   1.44 $ \\
110422   & $ 1.7700 $  &  $ 42.27  \pm   1.42 $ \\
080514B   & $ 1.8000 $  &  $ 44.45  \pm   1.46 $ \\
120326   & $ 1.8000 $  &  $ 44.09  \pm   1.43 $ \\
090902B   & $ 1.8220 $  &  $ 42.82  \pm   1.49 $ \\
131011A   & $ 1.8740 $  &  $ 42.75  \pm   1.44 $ \\
140623A   & $ 1.9200 $  &  $ 46.08  \pm   1.56 $ \\
080319C   & $ 1.9500 $  &  $ 45.30  \pm   1.51 $ \\
170113A   & $ 1.9680 $  &  $ 46.54  \pm   1.55 $ \\
081008   & $ 1.9685 $  &  $ 44.20  \pm   1.44 $ \\
030226   & $ 1.9800 $  &  $ 44.00  \pm   1.45 $ \\
170705A   & $ 2.0100 $  &  $ 43.81  \pm   1.42 $ \\
130612   & $ 2.0100 $  &  $ 46.48  \pm   1.44 $ \\
161017A   & $ 2.0130 $  &  $ 46.06  \pm   1.45 $ \\
140620A   & $ 2.0400 $  &  $ 44.57  \pm   1.42 $ \\
081203A   & $ 2.0500 $  &  $ 45.13  \pm   1.65 $ \\
150403A   & $ 2.0600 $  &  $ 44.78  \pm   1.50 $ \\
000926   & $ 2.0700 $  &  $ 43.37  \pm   1.44 $ \\
080207   & $ 2.0858 $  &  $ 44.47  \pm   1.68 $ \\
070521   & $ 2.0865 $  &  $ 44.70  \pm   1.44 $ \\
150206A   & $ 2.0870 $  &  $ 43.75  \pm   1.45 $ \\
061222A   & $ 2.0880 $  &  $ 44.72  \pm   1.47 $ \\
130610   & $ 2.0900 $  &  $ 46.02  \pm   1.45 $ \\
100728B   & $ 2.1060 $  &  $ 46.27  \pm   1.43 $ \\
090926A   & $ 2.1062 $  &  $ 42.61  \pm   1.44 $ \\
011211   & $ 2.1400 $  &  $ 44.53  \pm   1.43 $ \\
071020   & $ 2.1450 $  &  $ 46.11  \pm   1.54 $ \\
050922C   & $ 2.1980 $  &  $ 45.65  \pm   1.50 $ \\
121128   & $ 2.2000 $  &  $ 44.30  \pm   1.42 $ \\
120624B   & $ 2.2000 $  &  $ 42.93  \pm   1.45 $ \\
080804   & $ 2.2045 $  &  $ 45.69  \pm   1.44 $ \\
110205   & $ 2.2200 $  &  $ 44.28  \pm   1.55 $ \\
180325A   & $ 2.2480 $  &  $ 45.14  \pm   1.47 $ \\
081221   & $ 2.2600 $  &  $ 43.21  \pm   1.41 $ \\
130505   & $ 2.2700 $  &  $ 45.26  \pm   1.49 $ \\
140629A   & $ 2.2750 $  &  $ 45.12  \pm   1.47 $ \\
060124   & $ 2.2960 $  &  $ 44.35  \pm   1.52 $ \\
021004   & $ 2.3000 $  &  $ 45.72  \pm   1.53 $ \\
151021A   & $ 2.3300 $  &  $ 43.02  \pm   1.44 $ \\
141028A   & $ 2.3300 $  &  $ 44.63  \pm   1.46 $ \\
110128A   & $ 2.3390 $  &  $ 49.02  \pm   1.50 $ \\
051109A   & $ 2.3460 $  &  $ 45.95  \pm   1.50 $ \\
131108A   & $ 2.4000 $  &  $ 44.50  \pm   1.45 $ \\
171222A   & $ 2.4090 $  &  $ 44.29  \pm   1.44 $ \\
060908   & $ 2.4300 $  &  $ 45.52  \pm   1.45 $ \\
080413   & $ 2.4330 $  &  $ 45.97  \pm   1.50 $ \\
090812   & $ 2.4520 $  &  $ 45.71  \pm   1.56 $ \\
120716A   & $ 2.4860 $  &  $ 45.99  \pm   1.43 $ \\
130518A   & $ 2.4900 $  &  $ 44.01  \pm   1.47 $ \\
081121   & $ 2.5120 $  &  $ 41.56  \pm   1.49 $ \\
170214A   & $ 2.5300 $  &  $ 43.65  \pm   1.49 $ \\
081118   & $ 2.5800 $  &  $ 45.04  \pm   1.45 $ \\
080721   & $ 2.5910 $  &  $ 44.56  \pm   1.50 $ \\
050820   & $ 2.6120 $  &  $ 44.44  \pm   1.49 $ \\
030429   & $ 2.6500 $  &  $ 45.60  \pm   1.46 $ \\
120811C   & $ 2.6700 $  &  $ 44.06  \pm   1.43 $ \\
080603B   & $ 2.6900 $  &  $ 45.34  \pm   1.46 $ \\
161023A   & $ 2.7080 $  &  $ 43.99  \pm   1.46 $ \\
060714   & $ 2.7110 $  &  $ 44.56  \pm   1.58 $ \\
140206A   & $ 2.7300 $  &  $ 44.52  \pm   1.42 $ \\
091029   & $ 2.7520 $  &  $ 45.07  \pm   1.47 $ \\
081222   & $ 2.7700 $  &  $ 44.70  \pm   1.43 $ \\
050603   & $ 2.8210 $  &  $ 45.16  \pm   1.46 $ \\
161014A   & $ 2.8230 $  &  $ 46.21  \pm   1.43 $ \\
110731   & $ 2.8300 $  &  $ 45.35  \pm   1.45 $ \\
111107   & $ 2.8900 $  &  $ 46.92  \pm   1.48 $ \\
050401   & $ 2.9000 $  &  $ 44.50  \pm   1.47 $ \\
141109A   & $ 2.9930 $  &  $ 45.33  \pm   1.50 $ \\
090715B   & $ 3.0000 $  &  $ 45.31  \pm   1.49 $ \\
080607   & $ 3.0360 $  &  $ 44.50  \pm   1.48 $ \\
081028   & $ 3.0380 $  &  $ 44.58  \pm   1.51 $ \\
060607A   & $ 3.0820 $  &  $ 46.36  \pm   1.50 $ \\
120922   & $ 3.1000 $  &  $ 43.35  \pm   1.43 $ \\
020124   & $ 3.2000 $  &  $ 45.03  \pm   1.48 $ \\
060526   & $ 3.2100 $  &  $ 45.67  \pm   1.55 $ \\
140423A   & $ 3.2600 $  &  $ 44.43  \pm   1.43 $ \\
140808A   & $ 3.2900 $  &  $ 46.54  \pm   1.42 $ \\
160629A   & $ 3.3320 $  &  $ 45.82  \pm   1.46 $ \\
080810   & $ 3.3500 $  &  $ 46.14  \pm   1.48 $ \\
061222B   & $ 3.3550 $  &  $ 45.30  \pm   1.45 $ \\
110818   & $ 3.3600 $  &  $ 46.38  \pm   1.48 $ \\
030323   & $ 3.3700 $  &  $ 46.92  \pm   1.56 $ \\
971214   & $ 3.4200 $  &  $ 45.98  \pm   1.46 $ \\
060707   & $ 3.4250 $  &  $ 46.27  \pm   1.43 $ \\
170405A   & $ 3.5100 $  &  $ 44.37  \pm   1.46 $ \\
110721A   & $ 3.5120 $  &  $ 47.13  \pm   1.62 $ \\
060115   & $ 3.5300 $  &  $ 46.23  \pm   1.43 $ \\
090323   & $ 3.5700 $  &  $ 44.28  \pm   1.49 $ \\
130514   & $ 3.6000 $  &  $ 44.77  \pm   1.48 $ \\
100704   & $ 3.6000 $  &  $ 46.47  \pm   1.46 $ \\
130408   & $ 3.7600 $  &  $ 46.41  \pm   1.47 $ \\
120802   & $ 3.8000 $  &  $ 45.66  \pm   1.49 $ \\
100413   & $ 3.9000 $  &  $ 46.23  \pm   1.54 $ \\
060210   & $ 3.9100 $  &  $ 45.52  \pm   1.62 $ \\
120909   & $ 3.9300 $  &  $ 46.00  \pm   1.48 $ \\
140419A   & $ 3.9560 $  &  $ 45.29  \pm   1.76 $ \\
131117A   & $ 4.0400 $  &  $ 47.77  \pm   1.45 $ \\
060206   & $ 4.0480 $  &  $ 47.39  \pm   1.44 $ \\
090516   & $ 4.1090 $  &  $ 45.70  \pm   1.55 $ \\
120712A   & $ 4.1745 $  &  $ 46.51  \pm   1.45 $ \\
080916C   & $ 4.3500 $  &  $ 46.46  \pm   1.61 $ \\
000131   & $ 4.5000 $  &  $ 44.85  \pm   1.55 $ \\
090205   & $ 4.6497 $  &  $ 48.33  \pm   1.59 $ \\
140518A   & $ 4.7070 $  &  $ 46.92  \pm   1.45 $ \\
111008   & $ 5.0000 $  &  $ 46.43  \pm   1.48 $ \\
060927   & $ 5.6000 $  &  $ 47.21  \pm   1.43 $ \\
130606   & $ 5.9100 $  &  $ 48.48  \pm   1.53 $ \\
050904   & $ 6.2900 $  &  $ 47.58  \pm   1.59 $ \\
140515A   & $ 6.3200 $  &  $ 48.19  \pm   1.54 $ \\
080913   & $ 6.6950 $  &  $ 48.77  \pm   1.60 $ \\
120923A   & $ 7.8000 $  &  $ 49.36  \pm   1.49 $ \\
090423   & $ 8.2000 $  &  $ 48.49  \pm   1.54 $ \\
\hline

\end{longtable}
\end{appendix}

\end{document}